# Tailoring Dielectric Relaxation and Charge Transport in Amorphous Ge-Te-Se Phase-Change Materials through Yttrium Incorporation

Surbhi Agarwal[1*], Jiří Vacík[1], D.K. Dwivedi[2]
[1]Department of Neutron and Ion Methods
Nuclear Physics Institute of Czech Academy of Science, 25068 Husinec-Rez, Czech Republic
[2]Photonics and Photovoltaic Research Lab
Department of Physics and Material Science
Madan Mohan Malaviya University of Technology, Gorakhpur – 273010, India
*Email: agarwal.surbhi@ujf.cas.cz

**Abstract**

The $Te_{(1-x)}(GeSe_{0.5})Y_x$ (x = 0.05, 0.10, 0.15) alloys were prepared by a melt-quenching method and characterized using dielectric spectroscopy. The dielectric parameters and electrical conductivity were evaluated to assess their potential for phase-change memory and other electronic applications. The temperature and frequency dependence of the dielectric constants, ac conductivity, and dielectric loss of prepared chalcogenide glassy alloys were investigated over the frequency range 1 Hz - 1 MHz, and the temperature range 302 - 422 K, below the glass transition temperature. Both dielectric loss and dielectric constant were found to decrease with increasing frequency and decreasing temperature. The observed dielectric loss change was interpreted using the Guintini model of dielectric dispersion, correlated barrier hopping (CBH) model, and orientation polarization. The Cole-Cole analysis confirmed a distribution of relaxation times and yielded activation energies consistent with the CBH conduction mechanism. The frequency dependence of the ac conductivity follows the power law ($A\omega^s$), where the frequency exponent ($s < 1$) decreases with increasing temperature, further supporting the CBH conduction mechanism. The incorporation of Y increases the density of localized tail states, resulting in enhanced ac conductivity. These findings suggest that the investigated materials are promising candidates for future phase-change memory and other electronic device applications.

## 1. Introduction

Phase-change memory (PCM) has emerged as one of the most promising candidates for next-generation non-volatile memory because of its high operating speed, excellent endurance, low power consumption, and superior data retention compared with conventional flash memory. The operation of PCM devices relies on the reversible phase transition between the amorphous and crystalline states of chalcogenide materials, which exhibit a large contrast in their electrical and optical properties. Owing to their fast switching characteristics, high electrical resistivity contrast, and good thermal stability, chalcogenide glasses based on sulfur, selenium, and tellurium have attracted considerable attention as active materials for PCM applications[1–5].

Among various chalcogenide systems, Ge-Te-Se alloys have been extensively investigated because their structural and electrical properties can be tailored through compositional modification. In particular, $GeTeSe_{0.5}$ has been identified as a promising candidate for memory applications owing to its favorable thermal stability and phase-transition characteristics[6–8]. The introduction of a glassy network also allows tuning the physical-chemical properties through modification of the optical properties[9–11]. Previous studies have demonstrated that the addition of selenium increases the crystallization temperature and improves the stability of the amorphous phase, while also affecting the switching characteristics of Ge-Te-Se alloys[12]. Consequently, controlling the composition provides an effective approach for optimizing the performance of PCM materials.

Further improvement of chalcogenide glasses can be achieved by incorporating suitable dopants into the glass network. Rare-earth elements, such as yttrium, are known to modify the local structure and electronic properties of chalcogenide glasses, thereby influencing dielectric polarization, charge transport, and thermal stability[12–14]. Because of its high dielectric permittivity, good chemical stability, and favorable electrical properties, yttrium is expected to influence the dielectric response and electrical conduction of Ge-Te-Se-based glasses.

Dielectric spectroscopy provides valuable insight into the polarization and charge-transport mechanisms of chalcogenide materials. The frequency and temperature dependence of the dielectric constant, dielectric loss, and ac conductivity can be used to identify dielectric relaxation processes, evaluate charge-carrier dynamics, and determine the conduction mechanism. Several theoretical models, including correlated barrier hopping (CBH), non-overlapping small polaron tunneling (NOSPT), quantum mechanical tunneling (QMT), and overlapping large polaron tunneling (OLPT), have been proposed to describe ac conduction in chalcogenide glasses. Among these, the CBH model has been successfully applied to a wide range of Ge-based chalcogenide systems and is therefore adopted in the present work[15–17].

Although the structural and phase-change properties of Ge-Te-Se alloys have been widely investigated, relatively little attention has been paid to the influence of yttrium incorporation on their dielectric relaxation and ac conduction behavior. A detailed understanding of these properties is essential for evaluating the suitability of these materials for PCM and related electronic applications.

Therefore, the aim of this work was to investigate the frequency- and temperature-dependent dielectric properties of $Te_{(1-x)}(GeSe_{0.5})Y_x$ (x = 0.05, 0.10, and 0.15) glassy alloys prepared by the melt-quenching technique. The dielectric constant, dielectric loss, and ac conductivity were analyzed to identify the dominant conduction mechanism using the correlated barrier hopping model and to evaluate the effect of yttrium incorporation on the electrical behavior of these materials.

## 2. Experimental

$Te_{(1-x)}(GeSe_{0.5})Y_x$ (x = 0.05, 0.10, and 0.15) chalcogenide glasses were synthesized using the conventional melt-quenching technique (Fig. 1). High-purity elemental Ge, Te, Se, and Y (purity > 99.999%) were weighed according to the desired atomic compositions using an analytical balance (Yale LIBROR Model AEG 120). The elements were loaded into cleaned quartz ampoules, which were evacuated to a residual pressure of approximately $10^{-5}$ Torr and then sealed under vacuum.

The sealed ampoules were heated to 1000 °C and maintained at this temperature for 10-12 h to ensure complete melting and homogenization of the constituent elements. During melting, the ampoules were continuously rocked, while a ceramic rod placed inside each ampoule further promoted melt homogenization. After the heat treatment, the molten alloys were rapidly quenched in ice water to obtain the amorphous phase. Following cooling, the bulk glass samples were removed from the quartz ampoules.

For dielectric measurements, cylindrical pellets (10 mm in diameter and thickness d of ~ 1.2 mm) were prepared by uniaxially compressing finely ground powders under a load exceeding 5 tons using a hydraulic press.

Dielectric measurements were performed using a Solartron SI 1260 Impedance/Gain-Phase Analyzer over the frequency range of 1 Hz -1 MHz and the temperature range of 302 - 422 K. All measurements were carried out under a vacuum of approximately $10^{-3}$ Torr. The sample temperature was monitored using a copper-constantan thermocouple positioned close to the sample surface.

The impedance ($Z$), parallel capacitance ($Cd$), dissipation factor (tan δ), and phase angle ($\varphi$) were recorded over the entire frequency and temperature ranges. The dielectric loss angle was calculated from the relationship $\delta = 90° - \varphi$. The dielectric constant ($\varepsilon_1$) and dielectric loss ($\varepsilon_2$) were calculated from the measured capacitance and dissipation factor using the standard relationships[14,18–20],

$$\varepsilon_1 = Cd/\varepsilon_0 A \tag{1}$$

$$\varepsilon_2 = \varepsilon_1 \tan \delta, \tag{2}$$

where $\varepsilon_0$ is the permittivity of free space and $A$ is the electrode area.

The ac conductivity ($\sigma_{ac}$) was determined from

$$\sigma_{ac}(\omega) = \omega\varepsilon_0\varepsilon_2 = 2\pi f\varepsilon_2\varepsilon_1 \tan \delta, \tag{3}$$

where $f$ is the applied frequency.

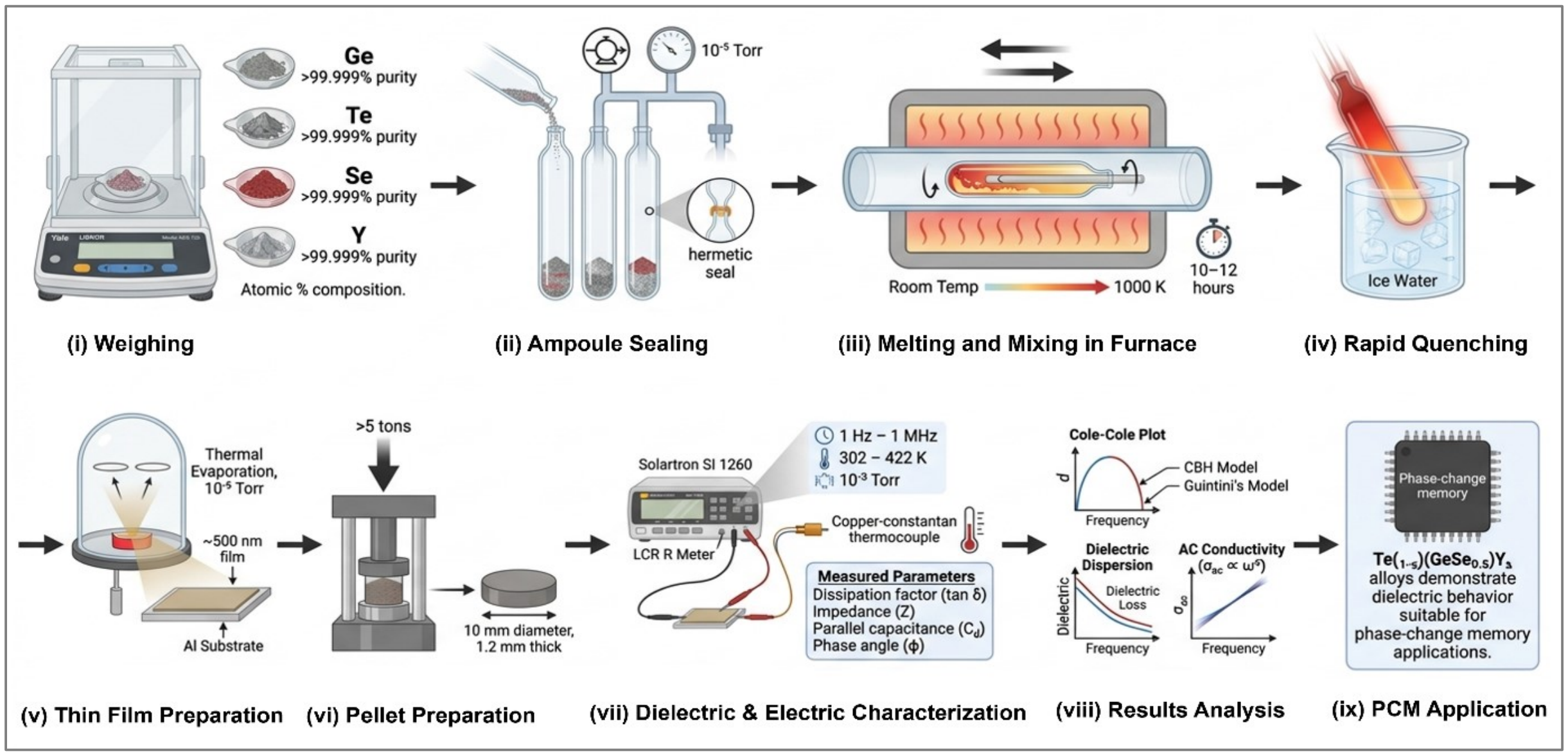

**Fig 1**. Schematic of melt-quench synthesis, sample preparation, and dielectric characterisation of $Te_{(1-x)}(GeSe_{0.5}Y_x)$ chalcogenide alloys for phase-change memory applications.

## 3. Theoretical background

Dielectric spectroscopy is a powerful technique for investigating polarization and charge-transport mechanisms in disordered semiconductors and chalcogenide glasses. The dielectric response of a material is described by the complex dielectric constant, $\varepsilon^* = \varepsilon_1 - i\varepsilon_2$, where $\varepsilon_1$ represents the dielectric constant and $i\varepsilon_2$ is the imaginary part of dielectric loss. The dielectric constant reflects the ability of a material to store electrical energy, whereas the dielectric loss characterizes energy dissipation under an alternating electric field. Both quantities depend on the frequency and temperature and are strongly influenced by the structural disorder, bonding configuration, and localized defect states present in chalcogenide glasses.

The dielectric relaxation behavior of chalcogenide glasses can be described by the model proposed by Giuntini *et al.* [2], which attributes dielectric loss to the hopping of charge carriers between localized defect states. According to this model, the dielectric loss is expressed by the following equation presented in [21]:

$$\varepsilon_2 = (\varepsilon_o - \varepsilon_\infty) 2\pi^2 N (ne^2/\varepsilon_o)^3 kT \tau_o^m W_m^{-4} \omega^m \tag{4}$$

where $\varepsilon_o$ and $\varepsilon_\infty$ are the static and high-frequency dielectric constants, respectively, ($N$) is the density of localized states, $n$ is the number of hopping electrons, $\omega$ is the angular frequency, and $W_m$ is the maximum barrier height.

At a fixed temperature, the dielectric loss follows the power-law relation[22]

$$\varepsilon_2 = A\omega^m \tag{5}$$

where the frequency exponent m is given by

$$m = -4kT/W_m \tag{6}$$

The value of m can be obtained from the slope of the $log\varepsilon_2$ versus $log(\omega)$ plot.

The total electrical conductivity is expressed as the sum of the *dc* and *ac* contributions,

$$\sigma_t(\omega) = \sigma_{ac}(\omega) + \sigma_{dc}(\omega) \tag{7}$$

where the *ac* conductivity obeys Jonscher's universal power law[23],

$$\sigma_{ac}(\omega) = A\omega^s \tag{8}$$

with *A* being a temperature-dependent constant and *s* the frequency exponent.

Several theoretical models have been proposed to explain ac conduction in amorphous semiconductors, including quantum mechanical tunneling (QMT), overlapping large polaron tunneling (OLPT), non-overlapping small polaron tunneling (NSPT), and correlated barrier hopping (CBH). Among these, the CBH model has been shown to provide an excellent description of charge transport in many chalcogenide glasses and is therefore adopted in the present work.

According to the CBH model, ac conduction originates from the hopping of electrons between oppositely charged localized defect states $D^+$ and $D^-$ over a potential barrier[16,24–26]. In the high-temperature limit ($\Delta_o \ll k_B T$), the ac conductivity is expressed as

$$\sigma_{ac}(\omega) = \frac{\pi^3}{24}\varepsilon_1\varepsilon_o N^2 \omega R_\omega^6 \tag{9}$$

where $N$ is the concentration of localized states and $R_\omega$ is the hopping distance,

$$R_\omega = e^2/\pi\varepsilon_1\varepsilon_o[W_m + k_B T ln(\omega T_o)] \tag{10}$$

For the low-temperature regime, Long[27,28] proposed the single-electron hopping expression

$$\sigma_{ac}(\omega) = \frac{\pi^3}{24}\frac{N^2}{\Delta_o^2}(k_BT)^2\varepsilon_1\varepsilon_o\omega R_\omega^6 \quad (11)$$

The density of localized states at the Fermi level $N(E_F)$ can be estimated using the Austin-Mott relation,

$$\sigma_{ac}(\omega) = (1/3)\pi e^2 k_B T N^2(E_F)\omega\beta^{-5}[\ln(v)_{ph}/\omega]^4 \quad (12)$$

where β is the decay constant of the electronic wave function and $(v)_{ph}$ is the phonon frequency $[(v)_{ph} = 10^4$ Hz and $\beta^{-1} = 10A^o]$.

Within the CBH model, the frequency exponent is given by

$$s = 1 - 6k_BT/[W_m + k_BTln(\omega\tau_o)] \quad (13)$$

which simplifies to

$$s = 1 - 6k_BT/W_m \quad (14)$$

when ($W_m >> k_BT$). A characteristic feature of the CBH model is that the frequency exponent decreases with increasing temperature, approaching unity as the temperature tends to zero. This temperature dependence serves as an important criterion for identifying the dominant conduction mechanism in chalcogenide glasses.

The relaxation time associated with electron hopping over the potential barrier is given by

$$\tau_o\exp(W_m/k_BT) \quad (15)$$

where $\tau_o$ is the characteristic relaxation time, typically of the order of $10^{-12}$ *s*. The activation energy for ac conduction can be obtained from the Arrhenius relation

$$\sigma_{ac}(\omega) = \sigma_o\exp(\Delta E(\omega)/k_BT) \quad (16)$$

where $\sigma_o$ is the pre-exponential factor and $\Delta E(\omega)$ is the frequency-dependent activation energy determined from the slope of the $ln(\sigma_{ac})$ versus $(1000/T)$ plot.

## 4. Results and Discussion

### 4.1. Dielectric constant behavior under varying temperature and frequency

The frequency dependence of the dielectric constant $\varepsilon_1$ of $Te_{(1-x)}(GeSe_{0.5})Y_x$ (x = 0.05, 0.1, 0.15) glassy alloys is presented in Fig. 2(a). For all investigated compositions, the dielectric constant decreases with increasing frequency. This behaviour is characteristic of amorphous chalcogenide materials and is commonly attributed to the frequency-dependent polarization mechanisms operating within the glass network.

The dielectric response arises from the combined contributions of electronic, ionic, dipolar (orientational), and space-charge polarization. At high frequencies, only the rapid electronic and ionic polarization mechanisms can follow the alternating electric field, whereas dipolar and space-charge polarization gradually diminish because of their longer relaxation times. Consequently, the dielectric constant decreases with increasing frequency and approaches an approximately constant value at the highest measured frequencies. Since the present measurements were performed over the frequency range of 1 Hz - 1 MHz, the observed dielectric dispersion is dominated primarily by dipolar and space-charge polarization.

The reduction in the dielectric constant with increasing frequency is mainly associated with the inability of dipoles and localized charge carriers to follow the rapidly alternating electric field. At low frequencies, charge carriers accumulate at defect sites and interfaces, resulting in enhanced space-charge polarization and higher dielectric constant values. As the frequency increases, this contribution progressively decreases, leading to the observed reduction in $\varepsilon_1$. Similar behaviour has been widely reported for chalcogenide glass systems[29–31].

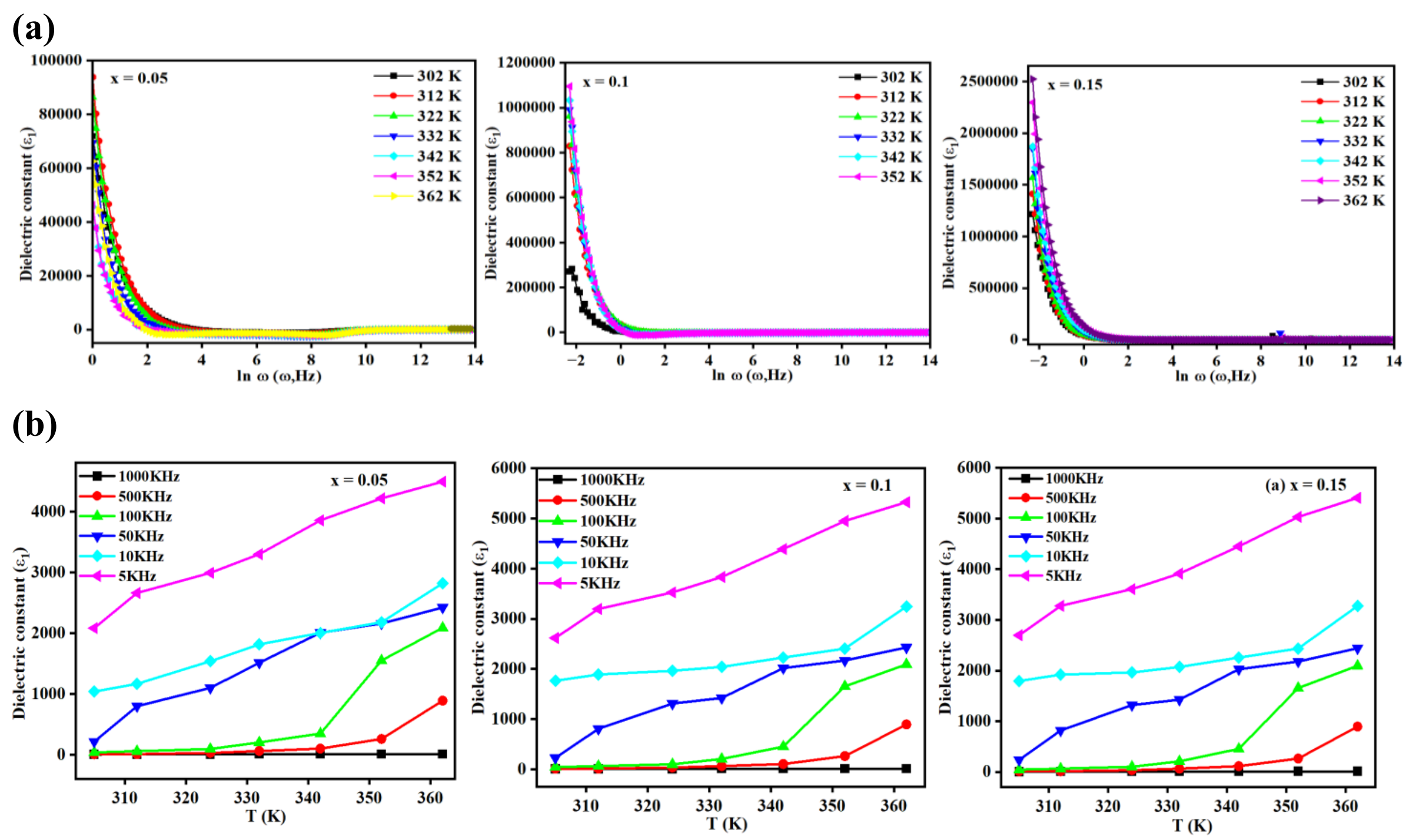

**Fig 2. (a)** Frequency dependency alloys, **(b)** Temperature dependency of dielectric constant of TGSY.

The temperature dependence of the dielectric constant is shown in Fig. 2(b). For all investigated frequencies, the dielectric constant increases with increasing temperature. This behaviour is attributed to enhanced dipolar orientation and thermally activated charge-carrier mobility, both of which contribute to increased polarization. As the temperature rises, charge carriers acquire sufficient thermal energy to overcome localized potential barriers, facilitating dipole reorientation and increasing the dielectric constant.

The temperature coefficient of permittivity (*TCP*), an important parameter for evaluating the thermal stability of dielectric materials, was calculated using

$$TCP = \frac{1}{\varepsilon_1}\frac{d\varepsilon_1}{dT} \tag{17}$$

The *TCP* values determined at 1 kHz and room temperature are summarized in Table 1. The obtained values are comparable with those reported previously for related chalcogenide glass systems[32–35].

## 4.2. Dielectric loss behaviour under varying temperature and frequency

The frequency dependence of the dielectric loss $\varepsilon_2$ for $Te_{(1-x)}(GeSe_{0.5})Y_x$ (x = 0.05, 0.10, and 0.15) glassy alloys is shown in Fig. 3(a). For all investigated compositions, the dielectric loss decreases with increasing frequency. This behavior is characteristic of amorphous chalcogenide materials and can be attributed to the reduced contribution of space-charge and dipolar polarization at higher frequencies. At low frequencies, charge carriers can follow the applied electric field and accumulate at defect sites and interfaces, resulting in enhanced dielectric loss. As the frequency increases, the ability of these carriers to respond to the alternating field decreases, leading to a reduction in $\varepsilon_2$.

The observed dielectric dispersion can be interpreted in terms of the model proposed by Mott and Davis[25,36–39], in which charge carriers hop between localized defect states under the influence of an alternating electric field. According to this model, electron hopping between donor $D^+$ and acceptor $D^-$ defect centers gives rise to dipolar entities that contribute to dielectric relaxation. The decrease in dielectric loss with increasing frequency is therefore associated with the limited response time of these hopping charge carriers.

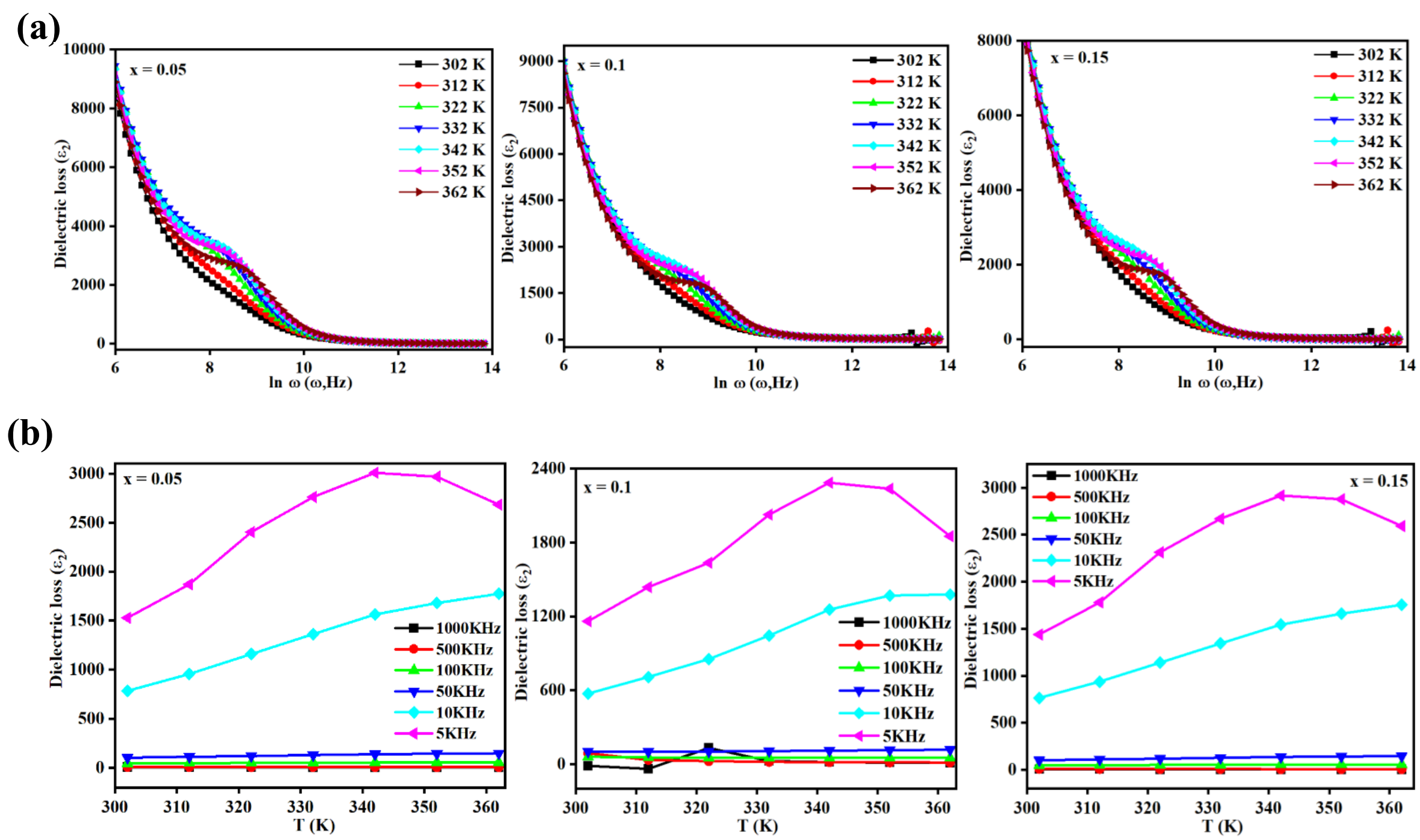

**Fig 3. (a)** Frequency dependency, **(b)** Temperature dependence of dielectric loss of TGSY alloys.

The temperature dependence of dielectric loss at different frequencies is presented in Fig. 3(b). It is evident that $\varepsilon_2$ increases with increasing temperature for all investigated frequencies. The increase in dielectric loss is attributed to the thermally activated nature of charge transport in chalcogenide glasses. As the temperature increases, charge carriers acquire sufficient energy to overcome localized potential barriers, resulting in enhanced hopping probability and increased dielectric loss. Similar behavior has been reported for other chalcogenide glass systems[21,40–42].

According to the correlated barrier hopping (*CBH*) model, dielectric relaxation is governed by charge carriers hopping over potential barriers separating localized states. The relaxation process is characterized by a maximum barrier height $W_m$, which determines the activation energy required for carrier hopping. The frequency dependence of dielectric loss can be analyzed using the Giuntini model, where the dielectric loss follows the relation (2) $\varepsilon_2=A\omega^m$ , and the exponent m is obtained from the slope of the $ln\varepsilon_2$ versus $ln(\omega)$ plots. The temperature dependence of ($m$) is presented in Fig. 4 and is discussed in terms of the CBH conduction mechanism.

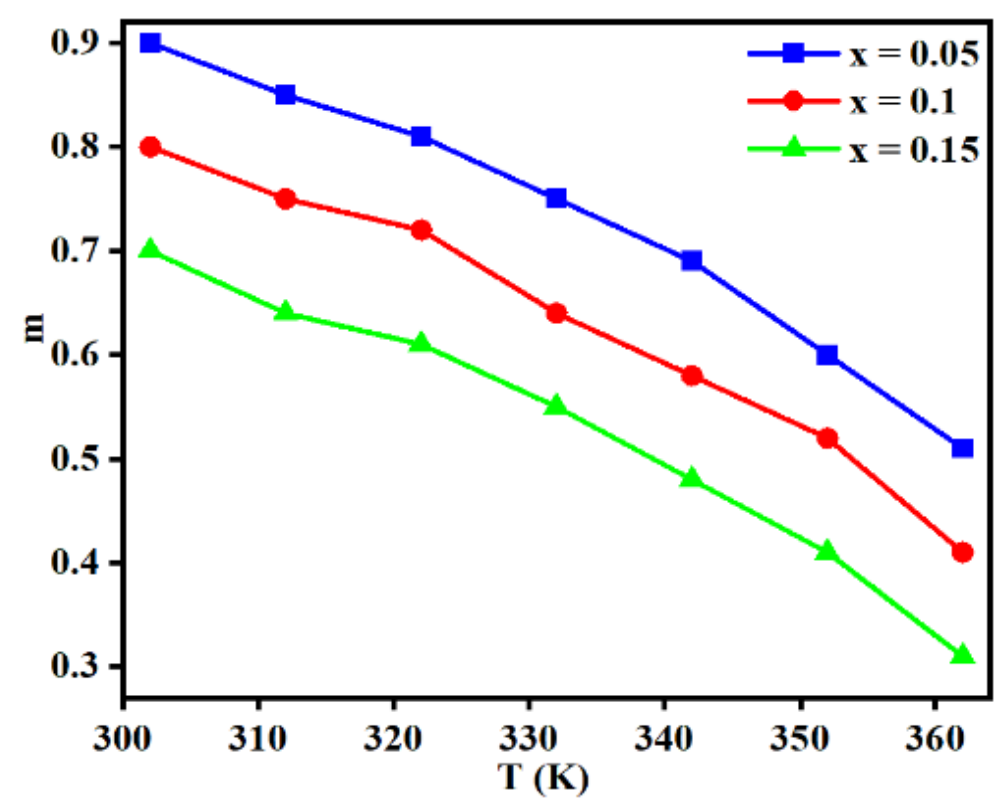

**Fig 4.** Variation of the dielectric loss frequency exponent m with temperature in TGSY alloys.

### 4.3. Evaluation of distribution parameter (α) and relaxation time (τ)

### 4.3 Cole-Cole analysis

Cole–Cole plots were employed to investigate the dielectric relaxation behavior of the $Te_{(1-x)}(GeSe_{0.5})Y_x$ (x = 0.05, 0.10, and 0.15) glassy alloys. For an ideal Debye dielectric, the Cole-Cole plot forms a perfect semicircle, whereas depressed semicircular arcs indicate a distribution of relaxation times associated with non-Debye relaxation. The characteristic dielectric parameters, including the static dielectric constant $\varepsilon_s$, optical dielectric constant $\varepsilon_\infty$, relaxation time $\tau_0$, distribution parameter $\alpha$, and resistive components $R_s$ and $R_\infty$, were determined from the Cole-Cole plots and are summarized in Table 2.

The maximum of the semicircular arc occurs when

$$\omega\tau = 1, \text{ or equivalently, } \omega = 1/RC \tag{18}$$

allowing the corresponding equivalent electrical circuit parameters to be determined. The macroscopic relaxation time $\tau_0$ was calculated from

$$u/v = (\omega\tau_o)^{1-\alpha} \tag{19}$$

where $u$ and $v$ represent the distances from the experimental point to $\varepsilon_\infty$ and $\varepsilon_s$, respectively. The parameter $\alpha$ characterizes the distribution of relaxation times. For an ideal Debye relaxation $\alpha = 0$, whereas $0 < \alpha < 1$ indicates a distribution of relaxation times arising from structural disorder within the glass network.

The molecular relaxation time $\tau$ was calculated using

$$\tau = [(2\varepsilon_s + \varepsilon_\infty)/3\varepsilon_s]\tau_o \quad (20)$$

The Cole-Cole plots recorded at different temperatures are presented in Fig. 5, while the corresponding values of $\alpha$, $\tau_0$, $\varepsilon_s$, and $\varepsilon_\infty$ are listed in Table 2. All compositions exhibit depressed semicircular arcs with their centers located below the real axis, confirming non-Debye dielectric relaxation. This behavior reflects the structural disorder, compositional fluctuations, localized defect states, and broad distribution of relaxation times that are characteristic of amorphous chalcogenide glasses[43,44].

Increasing the Y content systematically modifies the impedance response. For the (x = 0.05) composition, two distinguishable semicircular features are observed, indicating separate contributions from the bulk material and the grain-boundary/electrode interface. Accordingly, the impedance response can be represented by two parallel RC elements connected in series. For the (x = 0.10) composition, these relaxation processes partially overlap, suggesting that the bulk and interfacial responses occur simultaneously while remaining distinguishable. In contrast, the impedance spectrum of the (x = 0.15) alloy is dominated by a single depressed semicircle, indicating that the bulk response becomes predominant and the contribution from grain-boundary or interfacial polarization is considerably reduced. This evolution suggests that yttrium incorporation enhances the electrical homogeneity of the glass network while promoting thermally activated charge transport through the bulk of the material.

The depressed nature of the semicircular arcs indicates that the dielectric relaxation deviates from ideal Debye behavior. Such non-Debye relaxation is commonly attributed to the structural disorder, compositional inhomogeneity, and distribution of localized defect states that are intrinsic to amorphous chalcogenide materials[45,46].

The temperature dependence of the molecular relaxation time is presented in Fig. 6. The relaxation time decreases systematically with increasing temperature, indicating that dielectric relaxation is a thermally activated process. As the thermal energy increases, charge carriers more readily overcome the potential barriers separating localized states, resulting in faster relaxation dynamics. Similar behavior has been reported for a variety of chalcogenide glass systems[19,20,46,47].

The temperature dependence of the relaxation time follows the Arrhenius relation

$$\tau = \tau_\infty exp(\Delta E_\tau/kT) \quad (21)$$

where $\tau_0$ is the pre-exponential relaxation time corresponding to the limiting high-temperature value and $\Delta E_\tau$ is the activation energy for dielectric relaxation. The calculated values of $\tau_0$ and $\Delta E_\tau$ for the investigated TGSY glasses are summarized in Table 2.

**Table 1.** Dielectric and electrical parameters of $Te_{(1-x)}(GeSe_{0.5})Y_x$ (x = 0.05, 0.1, 0.15) alloys

| Composition | $W_m$ $(meV)$ | $\Delta E_\tau$ $(eV)$ | $\tau_\infty \times 10^{-4}(s)$ | TCP $(K^{-1})$ | $E_{MN}$ (meV) | $\ln\sigma_{oo}$ $\sigma_{oo}(\Omega.m)^{-1}$ |
|---|---|---|---|---|---|---|
| $Te_{0.95}(GeSe_{0.5})Y_{0.05}$ | 29.32 | 5.2 | 12.3 | 0.11 | 24.6 | -6.1 |
| $Te_{0.9}(GeSe_{0.5})Y_{0.1}$ | 24.15 | 4.6 | 5.6 | 0.07 | 20.1 | -3.8 |
| $Te_{0.85}(GeSe_{0.5})Y_{0.15}$ | 18.09 | 3.9 | 2.9 | 0.02 | 17.3 | -1.9 |

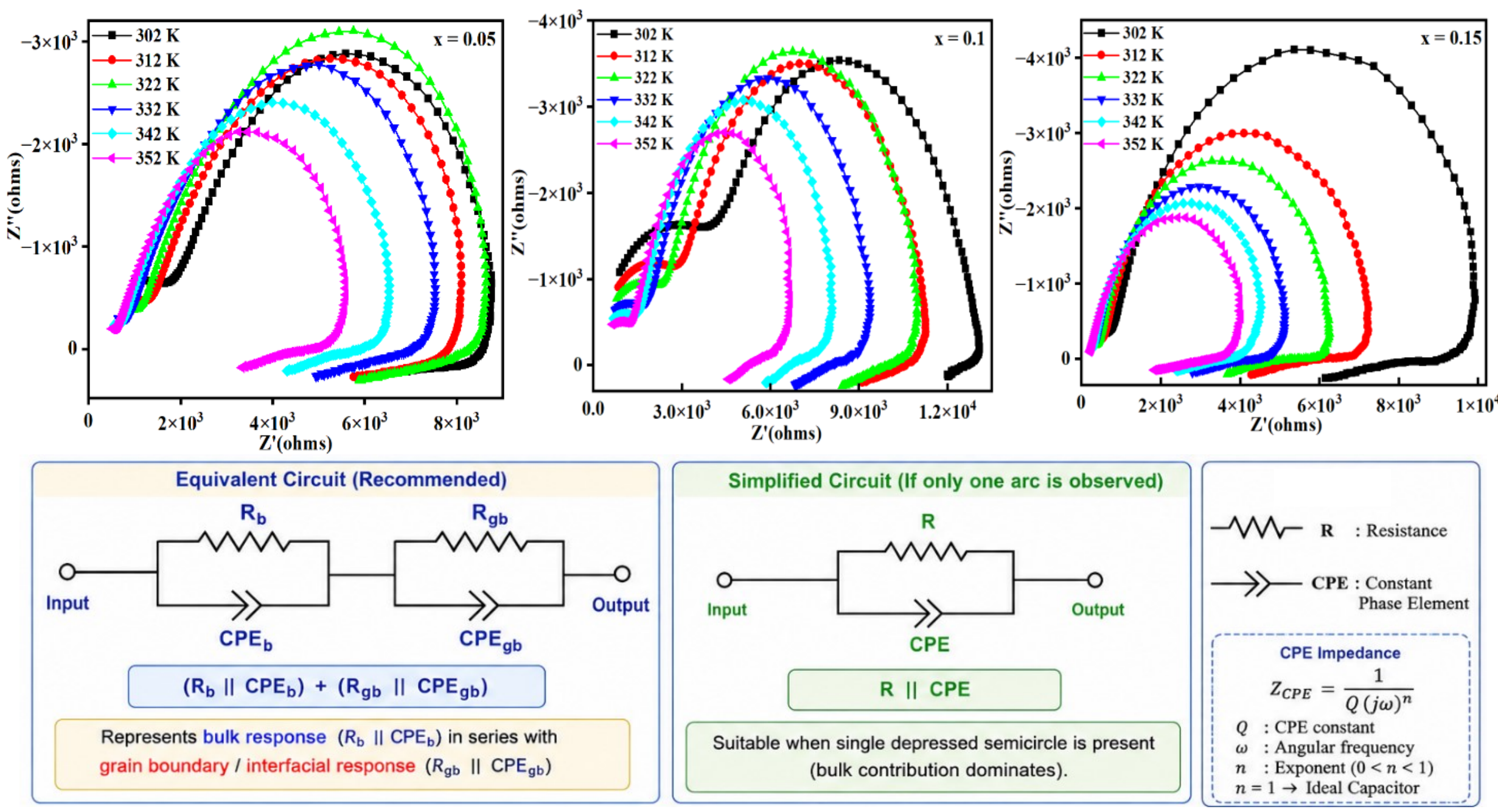

**Fig 5.** Cole-Cole diagram of TGSY alloys

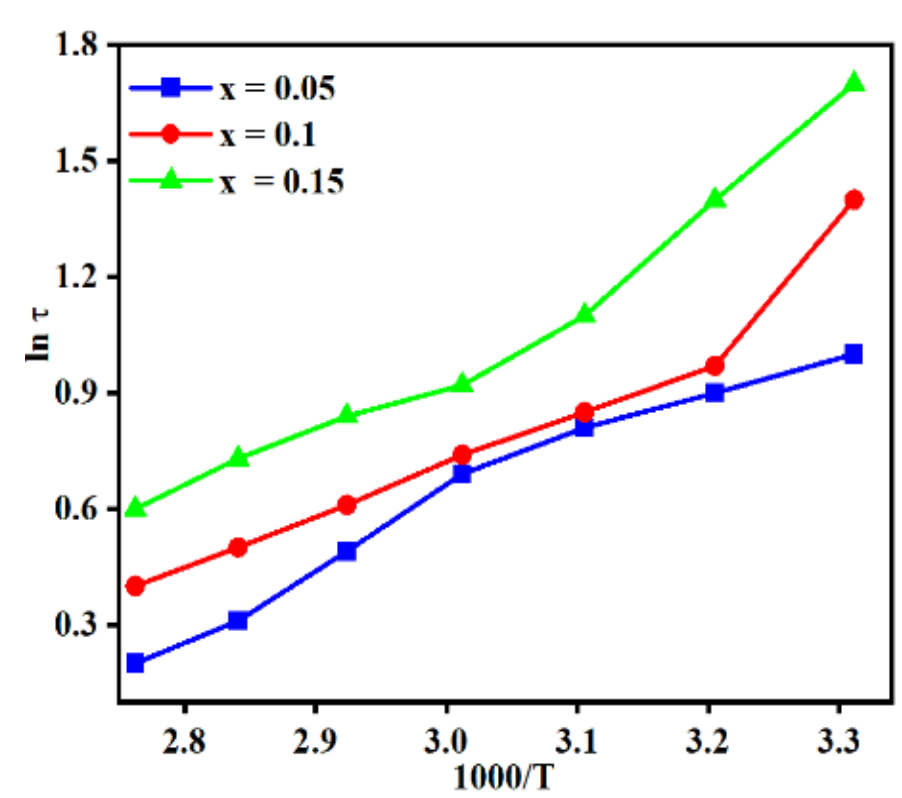

**Fig 6.** ln(τ) versus (1000/T) for TGSY alloy.

**Table 2.** The dielectric properties values of TGSY alloys**.**

| Composition | T(K) | $\varepsilon_s$ | $\varepsilon_\infty$ | $\tau_o \times 10^{-3}$(s) | $R_s \times 10^4$ | $R_\infty \times 10^4$ | $R \times 10^4$ (ohms) |
|---|---|---|---|---|---|---|---|
| $Te_{0.95}(GeSe_{0.5})Y_{0.05}$ | 302 | 20.1 | 16.4 | 4.26 | 15.9 | 0.13 | 53.9 |
| | 312 | 26 | 18.7 | 2.68 | 8.7 | 0.1 | 41.9 |
| | 322 | 29.6 | 20.2 | 1.57 | 3.3 | 0.08 | 16 |
| | 332 | 33.9 | 28.4 | 0.69 | 2.1 | 0.05 | 10.9 |
| | 342 | 36.2 | 33 | 0.42 | 0.9 | 0.01 | 6.6 |
| $Te_{0.9}(GeSe_{0.5})Y_{0.1}$ | 302 | 27.1 | 26.3 | 4.59 | 19.1 | 0.15 | 76.5 |
| | 312 | 34.8 | 28.3 | 2.69 | 9.1 | 0.11 | 38.1 |
| | 322 | 39.9 | 29.4 | 1.67 | 4.2 | 0.09 | 21.3 |
| | 332 | 43 | 35.1 | 0.71 | 2.5 | 0.08 | 8.5 |
| | 342 | 46.4 | 46.5 | 0.59 | 1.3 | 0.04 | 4 |
| $Te_{0.85}(GeSe_{0.5})Y_{0.15}$ | 302 | 33.7 | 35.5 | 4.71 | 21.5 | 0.18 | 3.2 |
| | 312 | 43.8 | 39.3 | 2.82 | 9.5 | 0.14 | 1.9 |
| | 322 | 50.1 | 39.8 | 1.91 | 5 | 0.1 | 0.6 |
| | 332 | 54.4 | 47.2 | 0.94 | 3.2 | 0.09 | 0.4 |
| | 342 | 56.6 | 53.7 | 0.62 | 1.8 | 0.07 | 0.2 |

## 4.3. Investigation of thermally activated ac conduction

### 4.3.1. Frequency dependence of ac conductivity ($\sigma_{ac}$)

The frequency dependence of the ac conductivity $\sigma_{ac}$ for the investigated $Te_{(1-x)}(GeSe_{0.5})Y_x$ (TGSY) glassy alloys is presented in Fig. 7(a). Since the *dc* conductivity is negligible within the investigated frequency range, the measured conductivity is dominated by the ac contribution. For all compositions, the ac conductivity increases with increasing frequency, which is characteristic of disordered semiconductors and amorphous chalcogenide glasses. The experimental data are well described by Jonscher's universal power law (Eq. (5)), $\sigma_{ac}(\omega) = A\omega^s$, where A is a temperature-dependent constant and s is the frequency exponent.

The temperature dependence of the frequency exponent s is shown in Fig. 7(b). The value of s remains below unity and decreases systematically with increasing temperature for all investigated compositions. Such behavior is a characteristic feature of the correlated barrier hopping (CBH) model and indicates that charge transport occurs predominantly through thermally activated hopping between localized defect states[46]. The decrease in s with temperature reflects the enhanced hopping probability of charge carriers as thermal energy assists them in overcoming the potential barriers separating neighboring localized states.

According to the CBH model proposed by Elliott[26,34,48–50], ac conduction arises from the hopping of charge carriers between oppositely charged localized defect states $D^+$ and $D^-$ separated by a Coulombic potential barrier. At elevated temperatures, thermally generated neutral defect states $D^0$ may also contribute to charge transport through single-polaron hopping, as suggested by Shimakawa[51], further enhancing the ac conductivity. The observed temperature dependence of the frequency exponent therefore confirms that the CBH model provides an appropriate description of the conduction mechanism in the investigated TGSY glassy alloys.

**Table 3.** The density of defect states *N(E<sub>F</sub>)* values for TGSY glasses at 10 kHz

| Temperature (K) | $N(E_F) \times 10^{22}$ ($m^{-3}$) | | |
|---|---|---|---|
| | $Te_{0.95}(GeSe_{0.5})\ Y_{0.05}$ | $Te_{0.9}\ (GeSe_{0.5})\ Y_{0.1}$ | $Te_{0.85}(GeSe_{0.5})\ Y_{0.15}$ |
| 302 | 13.18 | 21.14 | 31.39 |
| 312 | 15.28 | 24.82 | 35.57 |
| 324 | 18.06 | 27.93 | 38.88 |
| 332 | 24.24 | 34.48 | 45.02 |
| 342 | 30.97 | 42.48 | 51.91 |

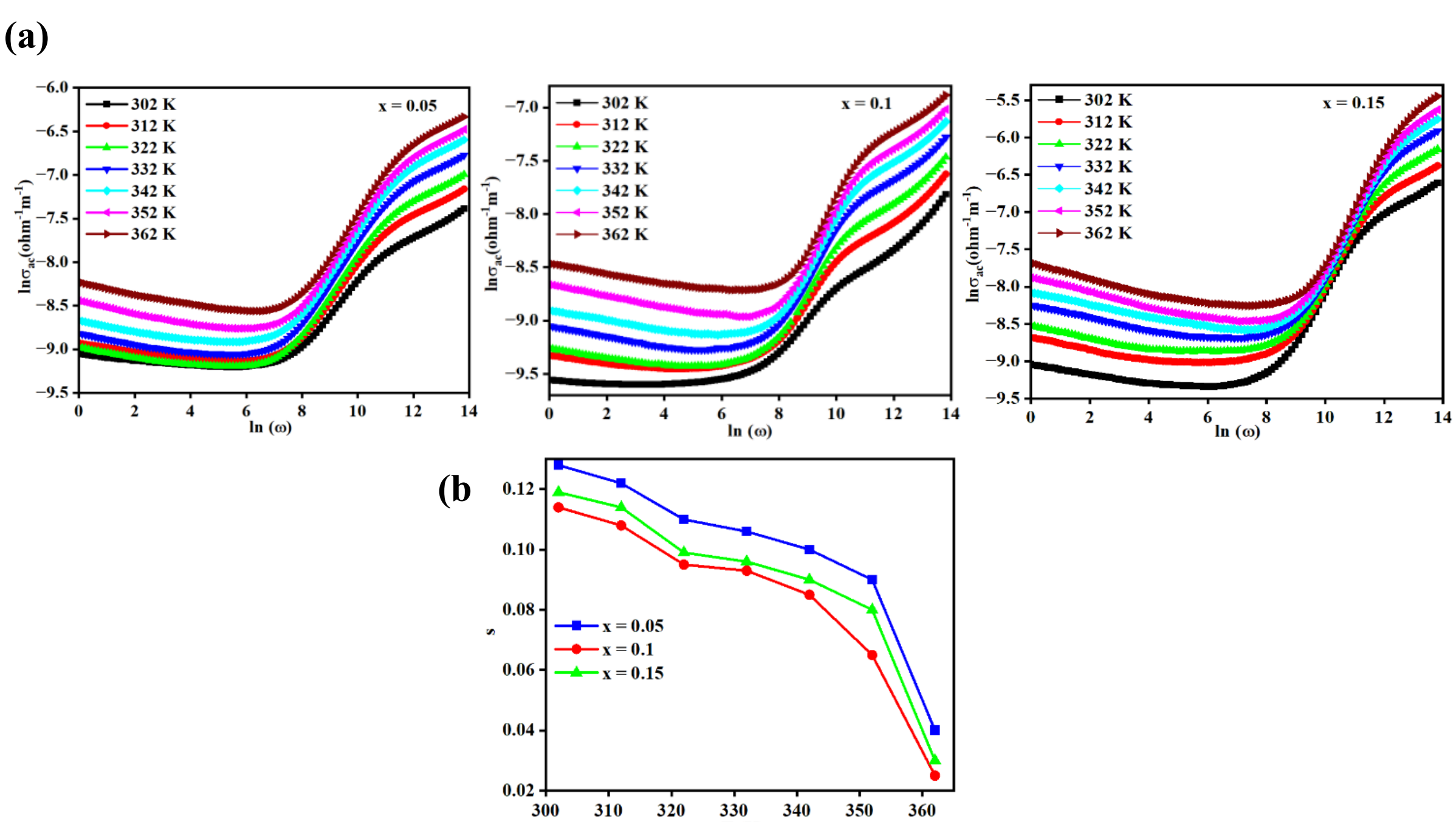

**Fig 7(a).** Variation in *ac* conductivity with frequency, **(b)** Dependency of frequency exponent *s* of *ac* conduction with temperature of TGSY alloys.

### 4.3.2. Temperature dependence of *ac* conductivity

The temperature dependence of the *ac* conductivity of the TGSY glassy alloys at different frequencies is presented in Fig. 8(a). For all investigated frequencies, the *ac* conductivity increases with increasing temperature, confirming that charge transport in these materials is a thermally activated process. The corresponding Arrhenius plots $ln(\sigma_{ac})$ versus *(1000/T)* exhibit an approximately linear behavior, indicating that the conduction mechanism is governed by a single dominant activation process involving hopping between localized states within the band gap.

The activation energy for ac conduction, $\Delta E(\omega)$, was determined from the slopes of the Arrhenius plots using Eq. (13). The dependence of $\Delta E(\omega)$ on frequency is shown in Fig. 8(b). The activation energy decreases systematically with increasing frequency, indicating that charge carriers require less thermal energy to hop between localized states under high-frequency excitation. This behavior is consistent with the correlated barrier hopping (*CBH*) model, in which the alternating electric field facilitates hopping by reducing the effective energy barrier for charge transport.

**(a)**

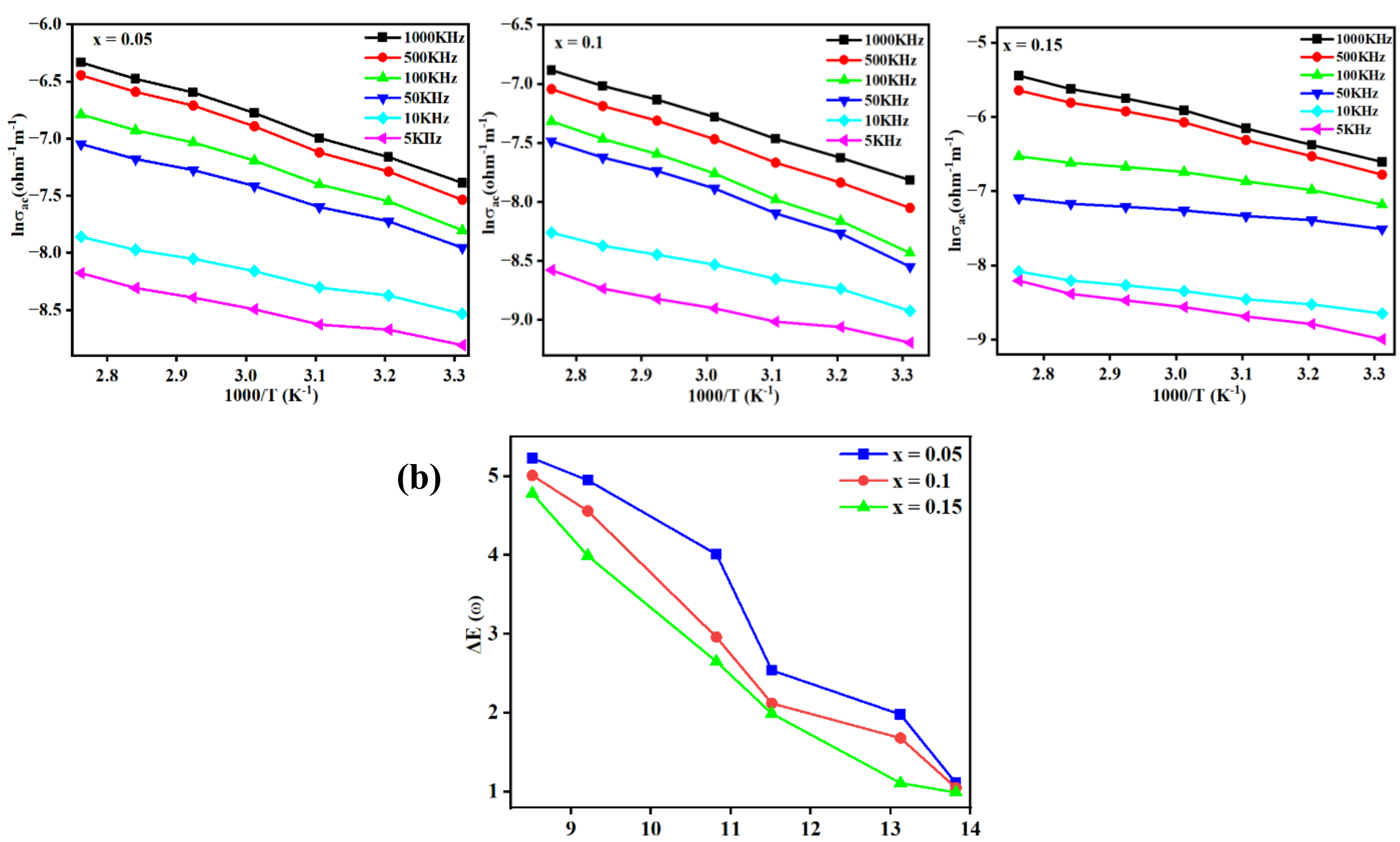

**Fig 8. (a)** Dependency of *ac* conductivity ($\sigma_{ac}$) with temperature, **(b)** The dependency of activation energy $\Delta E(\omega)$ of ac conduction against $ln(\omega)$ for TGSY alloys.

The values of the activation energy, together with the corresponding dielectric parameters, are summarized in Table 1. The observed frequency dependence of the activation energy is in good agreement with previous studies on amorphous chalcogenide glasses and further supports the applicability of the CBH model to the investigated TGSY alloys[40,52].

#### 4.3.3. Effect of yttrium incorporation on dielectric properties and ac conductivity

The effect of yttrium incorporation on the dielectric constant $\varepsilon_1$, dielectric loss $\varepsilon_2$, and *ac* conductivity $\sigma_{ac}(\omega)$ of the investigated TGSY glassy alloys is summarized in Fig. 9. A systematic increase in all three parameters is observed with increasing Y concentration, indicating that yttrium significantly modifies the dielectric response and electrical transport properties of the glass network.

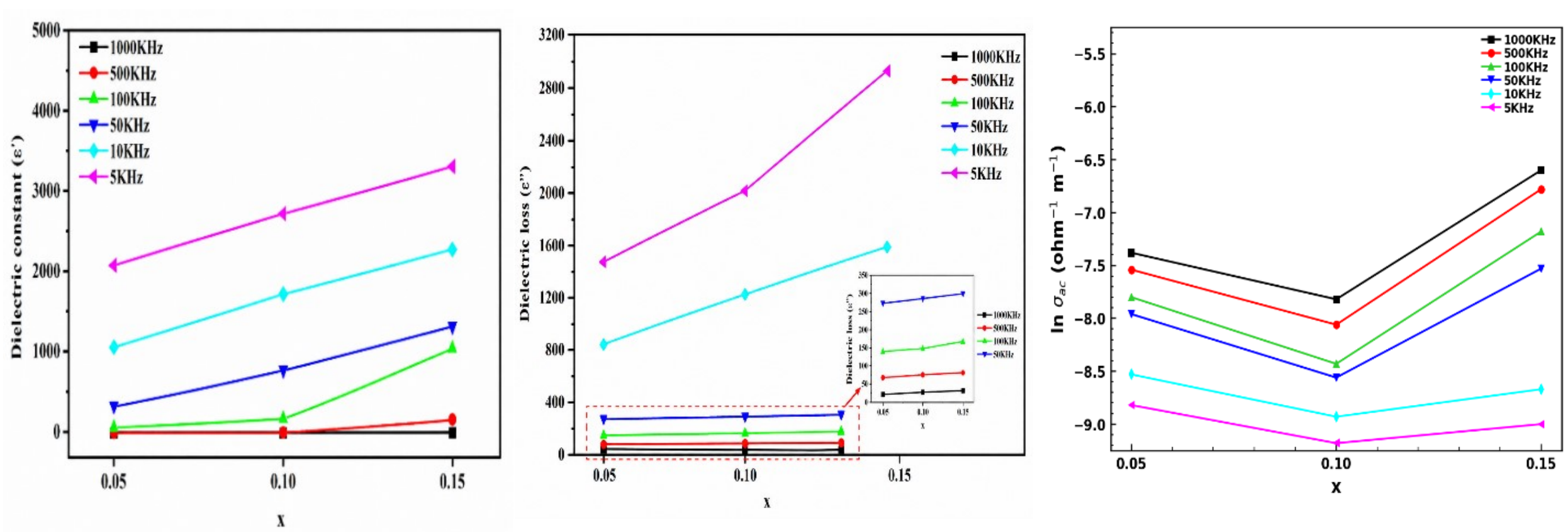

**Fig 9.** Dependence of dielectric constant $\varepsilon_1$, dielectric loss $\varepsilon_2$, and *ac* conductivity $\sigma_{ac}(\omega)$ on yttrium content.

The enhancement of the dielectric properties can be attributed to changes in the local atomic structure and electronic configuration induced by Y incorporation. Owing to its lower electronegativity and larger atomic radius compared with the constituent chalcogen atoms, yttrium alters the local bonding environment and promotes the formation of additional localized defect states within the mobility gap. These defect states act as hopping centers for charge carriers, increasing both the polarization response and the probability of thermally activated hopping conduction.

The increase in ac conductivity with increasing Y content is consistent with the correlated barrier hopping (CBH) model. Incorporation of yttrium increases the density of localized states near the band edges, thereby reducing the effective hopping barrier and facilitating charge transport between neighboring defect sites. Consequently, the dielectric constant, dielectric loss, and ac conductivity all increase as the Y concentration increases.

In addition to modifying the electronic structure, yttrium may also increase the cross-linking of the amorphous glass network, resulting in changes to the local structural arrangement and polarization behavior. Such structural modifications are expected to influence the distribution of localized states and contribute to the observed enhancement of the dielectric response. Similar effects of rare-earth incorporation on the electrical and dielectric properties of chalcogenide glasses have been reported previously[4,5,21,40].

Overall, the observed compositional dependence demonstrates that yttrium is an effective modifier of the dielectric and electrical properties of Ge-Te-Se glasses. The enhanced dielectric response and improved hopping conductivity suggest that controlled Y incorporation provides a viable route for tailoring chalcogenide glasses for phase-change memory and other electronic applications.

#### 4.3.4. Electrical impedance analysis

The frequency dependence of the real part of the impedance $Z'$ for the investigated TGSY glassy alloys is presented in Fig. 10(a). For all compositions, $Z'$ decreases with increasing frequency and temperature, indicating a progressive reduction in the overall impedance of the material. The higher impedance observed at low frequencies is associated with charge accumulation and interfacial polarization, whereas the continuous decrease in $Z'$ at higher frequencies reflects the enhanced mobility of charge carriers and the corresponding increase in *ac* conductivity.

The variation of the imaginary impedance $Z''$ with frequency at different temperatures is shown in Fig. 10(b). All compositions exhibit broad relaxation peaks that shift systematically toward higher frequencies with increasing temperature, demonstrating that the dielectric relaxation process is thermally activated. Simultaneously, the magnitude of $Z''$ decreases as the temperature increases, indicating a reduction in the bulk resistance and a shortening of the relaxation time. Furthermore, the relaxation peaks become broader with increasing Y concentration, suggesting an increasing distribution of relaxation times associated with the structural disorder characteristic of amorphous chalcogenide glasses.

The observed shift of the relaxation peaks toward higher frequencies confirms that charge carriers require progressively shorter times to respond to the alternating electric field as the temperature increases. This behavior is consistent with the thermally activated hopping mechanism described by the correlated barrier hopping (CBH) model.

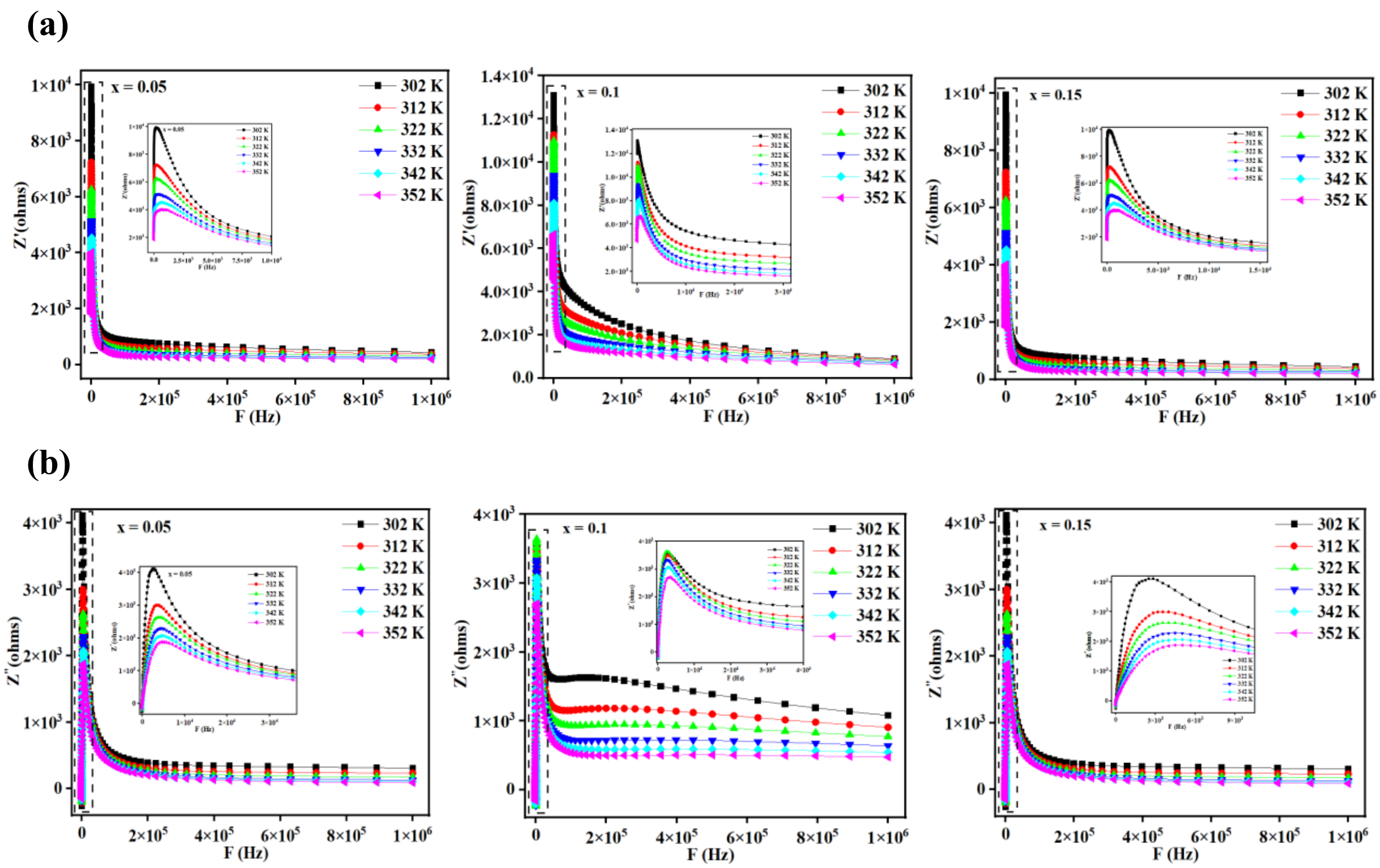

**Fig 10. (a)** Dependence of the real part $Z'$ of impedance, **(b)** variation of the imaginary part $Z''$ of impedance as a function of frequency at distinct temperatures for studied films.

**4.3.5. Phase-angle analysis**

The phase angle $\phi$ provides additional information about the dielectric relaxation and electrical response of the investigated materials. It is calculated from the real and imaginary components of the impedance according to

$$\emptyset = tan^{-1}\left(\frac{Z''}{Z'}\right) \tag{19}$$

where $Z'$ and $Z''$ are the real and imaginary components of the impedance, respectively.

The frequency dependence of the phase angle for the investigated TGSY glasses at different temperatures is presented in Fig. 11. At low frequencies, the phase angle remains close to zero, indicating that the electrical response is dominated by resistive conduction and interfacial polarization. With increasing frequency, the phase angle decreases and exhibits a broad minimum corresponding to dielectric relaxation associated with dipolar polarization and hopping of localized charge carriers.

The broad relaxation feature confirms that the investigated glasses exhibit non-Debye relaxation behavior arising from structural disorder and the distribution of localized relaxation times. As the temperature increases, the relaxation minimum shifts toward higher frequencies, indicating a decrease in the characteristic relaxation time and confirming the thermally activated nature of the charge-transport process.

The phase-angle spectra also reveal a clear compositional dependence. The alloy with (x = 0.10) exhibits a pronounced shoulder in the intermediate-frequency region, indicating the coexistence of bulk and interfacial relaxation processes. In contrast, the (x = 0.05) and (x = 0.15) compositions are characterized by a single broad relaxation feature, suggesting that one relaxation process dominates the impedance response. These observations are in good agreement with the Cole-Cole analysis and further support the existence of distributed relaxation processes in the investigated TGSY glassy alloys[30,53].

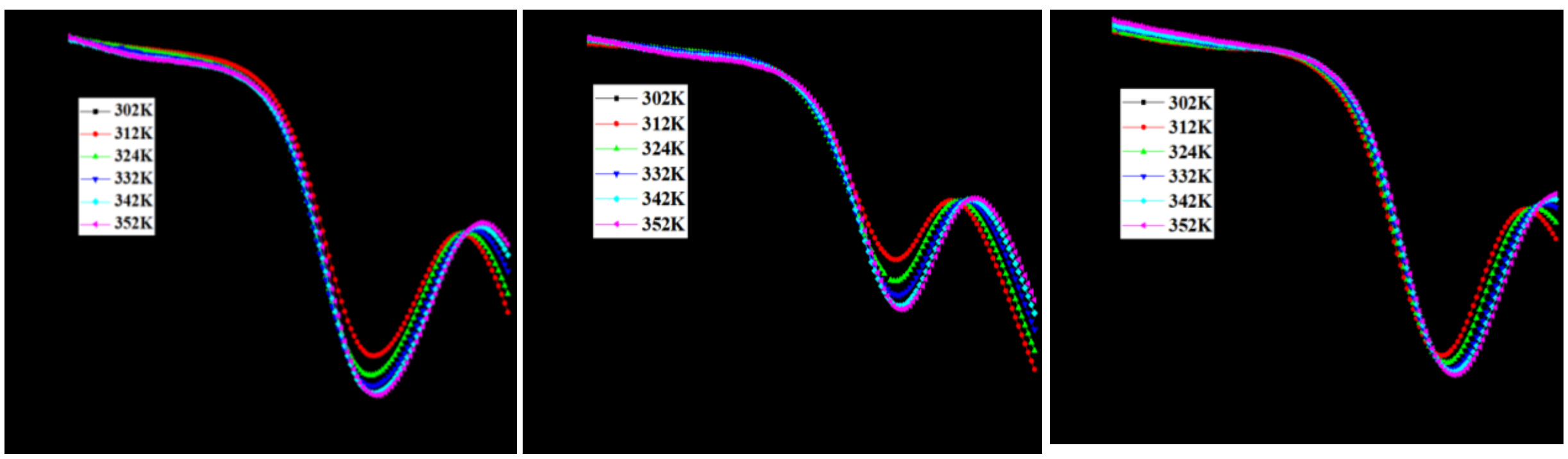

**Fig 11.** Phase angle variation with frequency at various temperatures

### 4.3.6. Polarization

The induced polarization $P$ of the investigated TGSY glassy alloys was calculated from the dielectric permittivity according to

$$P = \varepsilon_o(\varepsilon_r - 1)E \tag{20}$$

where $\varepsilon_o$ is the vacuum permittivity, $\varepsilon_r$ is the relative dielectric constant, and $E$ is the applied electric field. In the present study, the applied electric field was fixed at 400 V/m.

The temperature and frequency dependence of the induced polarization is presented in Fig. 12. For all investigated compositions, the polarization increases monotonically with increasing temperature and decreases with increasing frequency. The increase in polarization with temperature is attributed to enhanced thermally activated dipolar orientation together with increased hopping of localized charge carriers, both of which contribute to the overall dielectric response. In contrast, the reduction in polarization at higher frequencies results from the inability of slow polarization mechanisms, particularly dipolar and interfacial (Maxwell-Wagner) polarization, to follow the rapidly alternating electric field.

A clear compositional dependence is also observed. The alloy containing (x = 0.10) Y exhibits the highest induced polarization, suggesting that this composition provides the most favorable balance between structural disorder and charge-carrier polarization. The enhanced polarization is consistent with the increased dielectric constant and ac conductivity observed for this composition, indicating improved dielectric polarizability and charge-transport characteristics.

No polarization maximum was observed within the investigated temperature range (302 - 422 K), indicating that the dielectric response is governed primarily by thermally activated dielectric relaxation rather than by a ferroelectric phase transition. This observation is fully consistent with the non-Debye relaxation behavior identified from the Cole–Cole and impedance analyses[40,54,55].

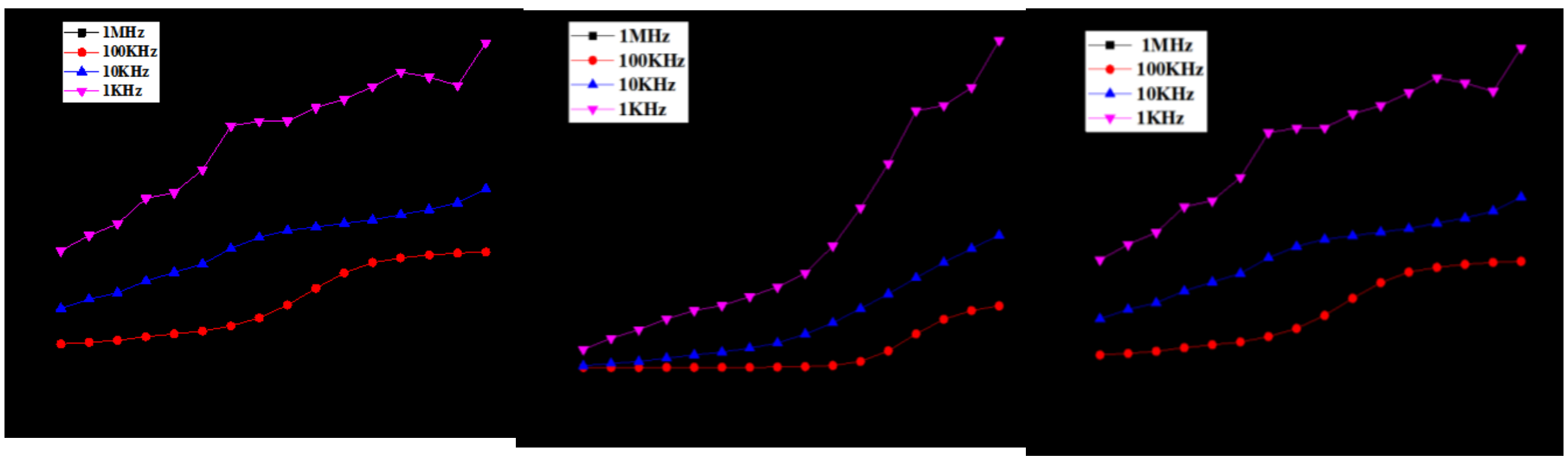

**Fig 12.** Variation of Polarization as a function of temperature at different frequencies.

## 5. Conclusions

The dielectric relaxation and ac conduction mechanisms of $Te_{(1-x)}(GeSe_{0.5})Y_x$ (x = 0.05, 0.10, and 0.15) glassy alloys were investigated over the frequency range of 1 Hz - 1 MHz and the temperature range of 302 - 422 K. The dielectric constant and dielectric loss decrease with increasing frequency and increase with increasing temperature, reflecting the combined effects of dipolar orientation and thermally activated charge-carrier hopping. The observed dielectric behavior is consistent with the Giuntini dielectric dispersion model, while the calculated maximum barrier heights and frequency exponent confirm that the correlated barrier hopping (CBH) model provides an appropriate description of the *ac* conduction mechanism.

Cole-Cole analysis revealed depressed semicircular arcs characteristic of non-Debye dielectric relaxation and enabled the determination of the relaxation parameters, including the distribution parameter, relaxation time, and activation energy. The relaxation time decreases with increasing temperature, confirming the thermally activated nature of the dielectric relaxation process. Impedance and phase-angle analyses further demonstrated that the electrical response is governed predominantly by intrinsic bulk relaxation associated with localized charge carriers rather than by electrode polarization.

The incorporation of yttrium enhances the dielectric constant, dielectric loss, and ac conductivity, indicating that Y modifies the electronic structure of the glass network by increasing the density of localized states available for hopping conduction. Among the investigated compositions, the alloy containing ($x = 0.10$) exhibited the most pronounced dielectric relaxation and polarization response, suggesting an optimum balance between structural disorder and charge transport.

Overall, the investigated Ge-Te-Se-Y glasses exhibit high dielectric polarizability, thermally activated hopping conduction, non-Debye dielectric relaxation, and relatively low dielectric loss at high frequencies. These characteristics are favorable for electronic materials requiring stable dielectric behavior and efficient charge transport, indicating that yttrium-modified Ge-Te-Se glasses are promising candidates for future phase-change memory applications.

**Applicability**

The present results demonstrate that yttrium incorporation provides an effective approach for tailoring the dielectric relaxation and electrical transport properties of Ge-Te-Se chalcogenide glasses. In addition to modifying the electrical response, the combined effects of Se and Y are expected to improve the thermal stability of the amorphous phase, suppress undesirable grain growth, and enable better control of crystallization behavior. These characteristics are particularly desirable for phase-change memory (PCM) applications, where an optimal balance between fast switching, high data retention, and long-term reliability is essential. Although further studies on crystallization kinetics, electrical switching behavior, and cycling endurance are required, the favorable dielectric and electrical properties observed in this work highlight the potential of Ge-Te-Se-Y alloys as promising candidates for next-generation phase-change memory devices and related electronic applications.

**Acknowledgements**

The authors acknowledge Nuclear Physics Institute of Czech Academy of Sciences, 25068 Husinec-Rez, Czech Republic for providing scientific infrastructure, technical assistance, and academic environment greatly contributed to the successful completion of this work. This work was supported by the Ministry of Education, Youth and Sports (MEYS) CR under the project OP JAK CZ.02.01.01/00/22_008/0004591. The authors also thank Prof. Prabhakar Singh, Professor at Department of Physics, IIT(BHU) Varanasi, India for access to characterization facilities and valuable scientific discussions. The authors sincerely thank all colleagues and staff members from both institutions for their valuable support and cooperation.

**Data Availability Statement**

Data will be made available on request.